\documentclass[useAMS,usegraphicx,usenatbib]{mn2e}

\newcommand{\dderiv}{\mathrm{d}}

\newcommand{\lmaxtune}{\ell_{\mathrm{max, tune}}}

\newcommand{\unit}[1]{\ensuremath{\,\mathrm{#1}}}

\newcommand{\Shalf}{\ensuremath{S_{1/2}}}

\def\lesssim{\mathrel{\hbox{\rlap{\hbox{\lower4pt\hbox{$\sim$}}}\hbox{$<$}}}}
\def\gtrsim{\mathrel{\hbox{\rlap{\hbox{\lower4pt\hbox{$\sim$}}}\hbox{$>$}}}}

\title{No large-angle correlations on the non-Galactic microwave sky}
\author[C.J. Copi, D. Huterer, D.J. Schwarz and G.D. Starkman]
{Craig J. Copi$^{1}$\thanks{E-mail: cjc5@cwru.edu},
  Dragan Huterer$^{2}$\thanks{E-mail: huterer@umich.edu},
  Dominik J. Schwarz$^{3}$\thanks{E-mail: dschwarz@physik.uni-bielefeld.de}
  and 
  Glenn D. Starkman$^{1}$\thanks{E-mail: glenn.starkman@case.edu}\\
  $^{1}$CERCA \& Department of Physics, Case Western Reserve University, Cleveland, 
  OH 44106-7079, USA\\
  $^{2}$Department of Physics, University of Michigan, 450 Church St, 
  Ann Arbor, MI 48109\\
  $^{3}$Fakult\"at f\"ur Physik, Universit\"at Bielefeld, Postfach 100131, 
  33501 Bielefeld, Germany}

\begin{document}

%\date{Accepted xxxx. Received xxxx; in original form xxxx}

\pagerange{\pageref{firstpage}--\pageref{lastpage}} \pubyear{2008}

\maketitle

\label{firstpage}

\begin{abstract}
  We investigate the angular two-point correlation function of temperature in
  the WMAP maps. Updating and extending earlier results, we confirm the lack
  of correlations outside the Galaxy on angular scales greater than about 60
  degrees at a level that would occur in 0.025 per cent of realizations of the
  concordance model.  This represents a dramatic increase in significance from
  the original observations by the COBE-DMR and a marked increase in
  significance from the first-year WMAP maps.  Given the rest of the reported
  angular power spectrum $C_\ell$, the lack of large-angle correlations that
  one infers outside the plane of the Galaxy requires covariance among the
  $C_\ell$ up to $\ell=5$.  Alternately, it requires both the unusually small
  (5 per cent of realizations) full-sky large-angle correlations, \textit{and} an
  unusual coincidence of alignment of the Galaxy with the pattern of
  cosmological fluctuations (less than 2 per cent of those 5 per cent).  We argue that
  unless there is some undiscovered systematic error in their collection or
  reduction, the data point towards a violation of statistical isotropy.  The
  near-vanishing of the large-angle correlations in the cut-sky maps, together
  with their disagreement with results inferred from full-sky maps, remain
  open problems, and are very difficult to understand within the concordance
  model.
\end{abstract}


\section{Introduction}

Over a decade ago, the Cosmic Background Explorer Differential Microwave
Radiometer (COBE-DMR) first reported a lack of large-angle correlations in
the two-point angular-correlation function, $\mathcal{C}(\theta)$, of the
cosmic microwave background (CMB) \citep{DMR4}.  This was confirmed by the
Wilkinson Microwave Anisotropy Probe (WMAP) team in their analysis of their
first year of data~\citep{Spergel2003}, and by us in the WMAP three-year
data~\citep{wmap123}.
Those findings have since been confirmed by \cite{Hajian:2007pi} and 
\cite{Bunn:2008zd}. Here, we present a more detailed analysis of the 
three-year and (for the first time) of the five-year WMAP data, 
confirming and strengthening our previous results. 

There is a common misconception that this lack of angular correlations is
equivalent to the low quadrupole in the two-point angular power spectrum,
which on its own does not have sufficient statistical significance to
challenge the canonical paradigm.  It is typically assumed both that the
angular power spectrum, $C_\ell$, and the two point angular correlation
function, $\mathcal{C} (\theta)$, contain the same information; and that
consequently studying one is as good as studying both.

Actually, the exact informational equivalence between $C_\ell$ and
$\mathcal{C} (\theta)$ holds only when the full sky is observed.
Statistically they are equivalent only when the sky is statistically
isotropic.  But even if $C_\ell$ and $\mathcal{C} (\theta)$ did contain the
same information, we are well aware that transforming between different
representations of the same information --- a time series and its Fourier
transform for example --- may make a real signal in the data easier or harder
to detect.  The Doppler peaks of the CMB, so clearly visible in the
$C_\ell$ representation are quite invisible in the two-point correlation
function.

The angular two-point function at the largest angular scales is our most
direct probe of the primordial seeds of structure formation (presumably
generated during cosmological inflation).  We expect that the large angular
scales are a direct probe of cosmological inflation, which predicts
statistically isotropic CMB temperature fluctuations generated by a
scale-invariant power spectrum of primordial quantum fluctuations. Without
inflation, at redshift $z \simeq 1100$ observed angular scales larger than
$1$ degree probe independent Hubble patches, while angular scales larger
than 60 degrees probe regions that are outside of causal contact until
$z\sim 1$. (More precisely, the post-inflation particle horizon subtends
$\theta\gtrsim 60\degr$ at $z\lesssim 4$ in the standard $\Lambda$CDM
model.) Therefore, the epoch of reionization and other secondary effects
(at $z > {\rm few}$) cannot modify the correlation function at these
scales. Any correlation on top of the primordial signal must be due to
local foregrounds (some contaminant at $z < \mathrm{few}$) or instrumental
systematic effects.

In this work we demonstrate that outside the region of the sky dominated by
our Galaxy, both of the CMB-dominated microwave bands --- V and W --- as well
as the Internal Linear Combinations (ILC) map synthesized from them as the
best map of the CMB, possess above $60\degr$ a level of two point angular
correlation higher than 99.975 percent of random realizations of the best-fit
$\Lambda$CDM model.  Indeed, above $60\degr$, $\mathcal{C} (\theta)$ is almost
entirely due to correlations involving points inside the Galaxy.

This level of statistical unlikelihood ($\mathcal{O}(10^{-4})$) should be
contrasted with what could be inferred from COBE ($\mathcal{O}(10^{-2})$).
This is a strong argument against
the criticism that its identification as an anomaly is \textit{a posteriori}.
It may have been \textit{a posteriori} for COBE, but its reidentification in
WMAP at dramatically increased statistical significance is precisely how one
goes about confirming that anomalies are actually present rather than
statistical accidents of an observation.

While the \textit{full-sky} map $\mathcal{C}
(\theta)$ itself has unexpectedly low large-angle correlations (occurring
only in 5 per cent of random realizations of the concordance model), we
find that what little correlation it does have is effectively ``hidden''
behind the Galaxy. In fact, we find that a random rotation of the Galactic
cut is as successful in masking the power only 2 per cent of the time,
in agreement with the previous claim that the little correlation above
$60\degr$ stems solely from two specific regions within the Galactic cut
covering just 9 per cent of the sky \citep{Hajian:2007pi}.  This further
underlies the striking lack of power outside the Galactic cut, and calls
into question cosmological uses of full-sky maps even for large angular
scale studies.

Finally, we demonstrate that the absence of large-angle correlations is
emphatically not a matter just of the low quadrupole.  Rather, given the
other measured multipoles, obtaining this little large-angle correlation
for the cut sky maps (i.e. the part outside the Galaxy) requires carefully
tuning $C_2$, $C_3$, $C_4$, and $C_5$.  There is also a strong indication
that it is not enough to find a model in which the theoretical $C_\ell$
yield a very small correlation function on large angular scales.  This is
because, even if the theoretical $C_\ell$ were to be set equal to those
that are inferred from the cut-sky $\mathcal{C}(\theta)$ --- so that the
expected $\mathcal{C}(\theta)$ nearly vanished above $60\degr$ --- an
actual realization of Gaussian-random statistically independent $a_{\ell
  m}$ with these ${C}_{\ell}$ would yield different observed
$\mathcal{C}_\ell$ because of cosmic variance.  $C(\theta>60\degr)$ would
then not be nearly so close to zero.  Thus getting $C(\theta>60\degr)$ to
vanish as it does seems to require covariance among the low-$\ell$
$C_\ell$, and thus among $a_{\ell m}$ of different $\ell$.  This is in
contradiction to the predictions of standard inflationary cosmological
theory.

One is therefore placed between a rock and a hard place.  If the WMAP ILC is a
reliable reconstruction of the full-sky CMB, then there is overwhelming
evidence (\citet{deOliveira2004,Eriksen_asym,Copi2004,Schwarz2004,
  lowl2,wmap123,Land2005a,Land2004a, Land2004b,Land2005b,Rakic:2007ve}; for a
review see \citet{Huterer_NewAst_review}) of extremely unlikely phase
alignments between (at least) the quadrupole and octopole and between these
multipoles and the geometry of the Solar System --- a violation of statistical
isotropy that happens by random chance in far less than 0.025 per cent of
random realizations of the standard cosmology.  If, on the other hand, the
part of the ILC (and band maps) inside the Galaxy are unreliable as
measurements of the true CMB, then the alignment of low-$\ell$ multipoles
cannot be readily tested, but the magnitude of the two-point angular
correlation function on large angular scales outside the Galaxy is smaller
than would be seen in all but a few of every 10,000 realizations.

We can only conclude that
(i) we don't live in a standard $\Lambda$CDM Universe with a standard
  inflationary early history;
(ii) we live in an extremely anomalous realization of that cosmology;
(iii) there is a major error in the observations of both COBE and WMAP; or
(iv) there is a major error in the reduction to maps performed by both
  COBE and WMAP.
Whichever of these is correct, inferences from the large-angle data 
about precise
values of the parameters of the standard cosmological model should be
regarded with particular skepticism.

Finally we note that there is no single test for statistical
\textit{anisotropy}.  There are countless ways of breaking statistical
isotropy, that is, of having $\langle a^*_{\ell m} a_{\ell' m'}\rangle \ne
\mathcal{C}_\ell \delta_{\ell \ell'} \delta_{m m'}$.  Any one of them can
be tested against the data but no single test can cover all possibilities.
Different tests will be sensitive to different ways of breaking statistical
isotropy.  Thus it is both a boon and a bane that there are multiple tests
with varying results (e.g.\ non-detections of violation of statistical
isotropy in \cite{Hajian_Souradeep_2006} and \cite{Dennis_Land}) discussed
in the literature.  Ideally these tests will lead to an understanding of
how statistical isotropy can be broken and may ultimately provide an
explanation of the source of the signatures seen in some tests and not in
others.  In the remainder of this paper we provide detailed discussion of
the tests we apply and the evidence and reasoning for the statements made
in the previous paragraph.

\section{Angular correlation function: preliminaries}
\label{sec:preliminaries}

The two-point correlation function of the observed CMB temperature
fluctuations
\begin{equation}
  \mathcal{C} (\theta) \equiv \overline{T (\hat e_1) T (\hat e_2)}_\theta,
  \label{eq:calCthetaavg}
\end{equation}
where the over-bar represents an average over all pairs of points on the
sky (or at least that portion of the sky being analyzed) that are separated
by an angle $\theta$.  On the one hand, we are interested in this quantity
as a partial characterization of the observations.  On the other hand, we
regard it as an (unbiased) estimator of the ensemble average of the same
quantity --- where the ensemble is of realizations of the sky in a
particular model cosmology.  

It is commonly thought that $\mathcal{C}(\theta)$ 
contains the same information as the angular power spectrum,
\begin{equation}
  C_\ell \equiv \frac{1}{2\ell+1} \sum_{m=-\ell}^\ell \left| a_{\ell m}
  \right|^2.
  \label{eq:Cl}
\end{equation}
(Here $a_{\ell m}$ are the coefficients of a spherical harmonic
decomposition of the temperature fluctuations on the sky.)  This is
because, \textit{for a full sky},
\begin{equation}
  \label{eq:Ctheta}
  \mathcal{C} (\theta) = 
  \frac1{4\pi} \sum_{\ell=0}^\infty (2\ell+1) C_\ell P_\ell (\cos\theta) .
\end{equation}
Again the $C_{\ell}$ are regarded as most interesting to us as unbiased
estimators of the ensemble average of $\left| a_{\ell m} \right|^2$.
Furthermore, the standard inflationary model predicts that the Universe
is statistically isotropic, so that the ensemble average of pairs 
of $a_{\ell m}$ are independent:
\begin{equation}
	\label{eq:SI}
	\left \langle a^\star_{\ell m} a_{\ell' m'}\right\rangle = 
          \mathcal{C}_{\ell} \delta_{\ell\ell'} \delta_{m m'} .
\end{equation}
Theoretically, the  $C_{\ell}$ therefore encode all of the information 
from the sky that has cosmological significance.

Actually, $C_{\ell}$ and $\mathcal{C}(\theta)$ only contain precisely the
same information for full-sky data.  Their analogues in the ensemble are
informationally equivalent only if, in the ensemble, the sky is statistically
isotropic.  This suggests that by measuring both $\mathcal{C} (\theta)$ and
$C_\ell$ we can probe the correctness of the assumption of statistical
isotropy of the Universe.  Statistical isotropy is a fundamental prediction
of generic inflationary models.

More importantly, it is well known that, while a function and its Fourier
transform posses exactly the same information differently organized, in
different circumstances one or the other may more clearly show an
interesting feature.  For example, a sharp delta-function spike in a time
series will merely contribute equally to all modes of the associated
Fourier series. It is precisely the same with the two-point
angular-correlation function and its Legendre-transform, the angular power
spectrum.  Thus, while our theory may suggest to us that it is easier to
analyze the angular power spectrum, prudence demands that we also consider
the properties of the angular correlation function, all the more so since
our actual measurements are done in ``angle-space'' not in
``$\ell$-space''.

In order to highlight these differences, we use the calligraphic 
symbol, $\mathcal{C}$, for objects operationally defined in 
``angle-space'' and the symbol, $C$, for quantities in ``$\ell$-space'';
e.g.~the Legendre transform of the two-point correlation function 
(\ref{eq:calCthetaavg}) is
\begin{equation}
  \mathcal{C}_\ell \equiv 2\pi \int_{-1}^{1} P_\ell(\cos \theta)
  \mathcal{C}(\theta) \dderiv (\cos \theta).
  \label{eq:calCl}
\end{equation}
Note that $\mathcal{C}_\ell$ can be negative --- in contrast to the
angular power spectrum $C_\ell$ as defined in (\ref{eq:Cl}). 

\begin{figure}
  \includegraphics[angle=-90,width=3.5in]{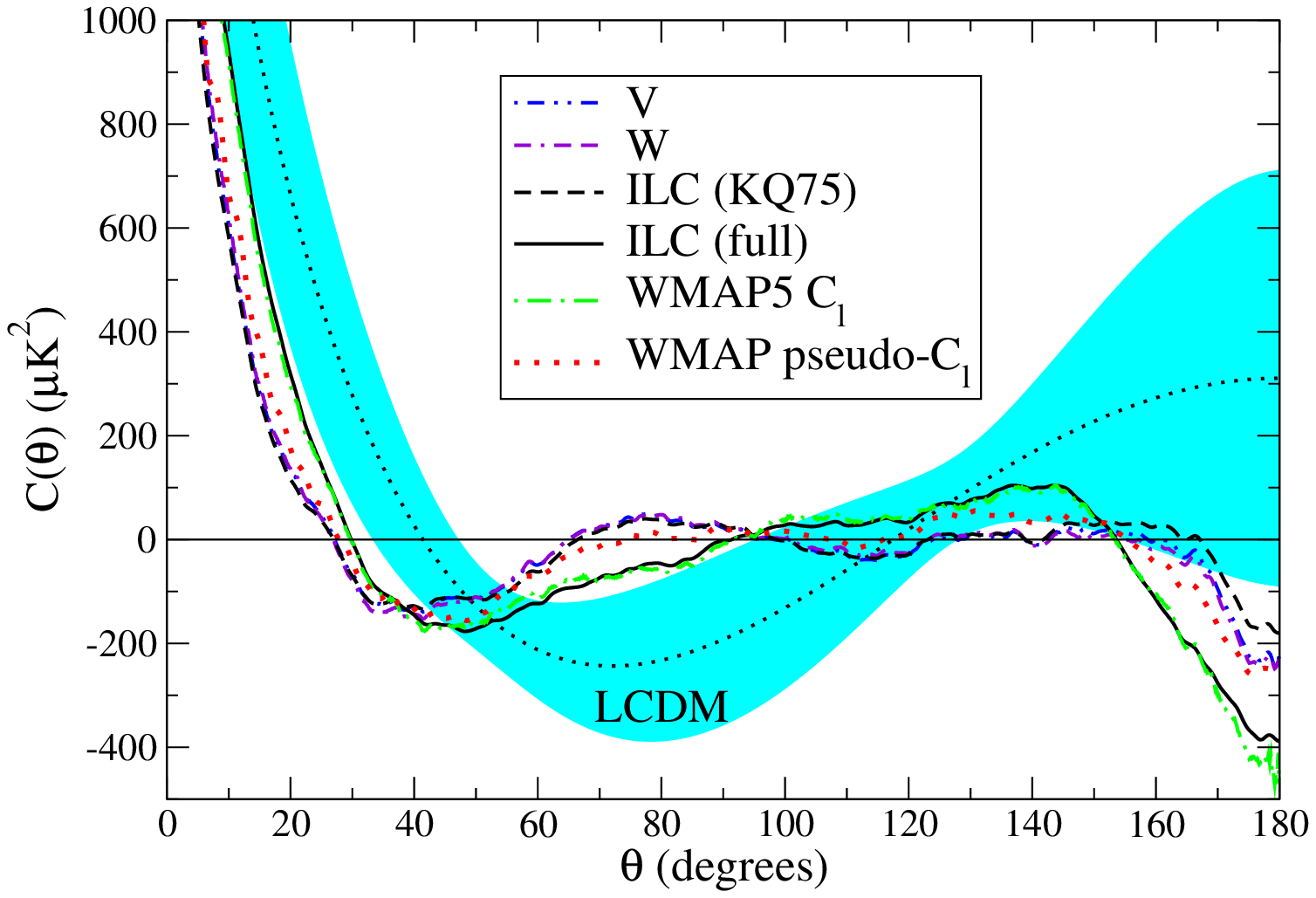}
  \caption{The two-point angular correlation function from the WMAP 5 year
    results.  Plotted are $\mathcal{C} (\theta)$ for maps with Doppler 
    quadrupole subtracted.  The V (dashed-dotted-dotted line), 
    W (dashed-dashed-dotted line), ILC (KQ75, dashed line) have had the 
    KQ75 mask applied.  The full-sky ILC result (solid line) is also shown.  
    Also plotted are $C (\theta)$ from the WMAP maximum likelihood 
    $C_\ell$ (dotted-dashed line), the WMAP pseudo-$C_\ell$ (dotted line) 
    and the best-fit $\Lambda$CDM $C_\ell$.  The shaded region is the one 
    sigma cosmic variance bound on the standard $\Lambda$CDM theory.}
  \label{fig:ctheta}
\end{figure}

The angular correlation functions in this work have been calculated using
SpICE \citep{SpICE} at \texttt{NSIDE=512} for data maps and at
\texttt{NSIDE=64} for the Monte Carlo studies.  The map average has been
subtracted in all cases.  The results are shown in Fig.~\ref{fig:ctheta}
for four different maps --- the ILC map, which covers the full sky, and
KQ75 cut-sky versions of the ILC, the V-band, and the W-band.  In the same
figure, we have plotted the Legendre transform of the angular power
spectrum (cf. equation \ref{eq:Ctheta}) calculated using both the
pseudo-$C_\ell$ method (essentially equation \ref{eq:Cl}) applied by the
WMAP team in their first-year analysis, and the maximum likelihood
estimates of the angular power spectrum as used by WMAP in the third- and
fifth-year analysis.
Finally we have plotted the
expected $C(\theta)$ for the best-fit $\Lambda$CDM, and, in blue,
the one-sigma cosmic-variance band around the best fit.

Three striking observations should be made about
$\mathcal{C}(\theta)$:
\begin{enumerate}
\item None of the observational angular correlation functions visually
  match the expectations from the theoretical
  model.
\item All of the cut-sky map curves are very similar to each other, and
  they are also very similar to the Legendre transform of the
  pseudo-$C_\ell$ estimate of the angular power spectrum.  Meanwhile the
  full-sky ILC $\mathcal{C}(\theta)$ and the Legendre transform of the MLE
  of the $C_\ell$ agree well with each other, but not with any of the
  others.
\item The most striking feature of the cut-sky (and pseudo-$C_\ell$)
  $\mathcal{C}(\theta)$, is that all of them are very nearly zero above
  about $60\degr$, except for some anti-correlation near $180\degr$. This
  is also true for the full-sky curves, but less so.
\end{enumerate}

In order to be more quantitative about these observations, we must adopt
some statistic that measures large-angle correlations.  This means that we
must identify some norm that measures the difference between two functions
over a range of angles.  Different choices of norm, or different choices
for the angular range, will give slightly different numerical results for
the improbability of the above observations; however, as we shall see, the
observations are so unlikely that we can be confident that reasonable
choices of the norm lead to similar results.

In their analysis of the first year data, the WMAP team defined the \Shalf\
statistic~\citep{Spergel2003}
\begin{equation}
  \label{eq:S12}
  \Shalf \equiv \int_{-1}^{1/2} \left[ \mathcal{C}(\theta)\right]^2 \dderiv
  (\cos\theta).
\end{equation}
While the choice of $1/2$ as the upper limit of the integral, and the
particular choice of a square norm were \textit{a posteriori}, they are
neither optimized nor particularly special.  In fact, this two-point
correlator is the most basic quantity to study and $\Shalf$ is probably the
simplest statistic that tests the total amount of correlations at large
angles.  Moreover, the absence of large-angle correlations was noted by the
COBE team (though without definition of a particular statistical measure),
and the choice of $\sim60\degr$ is clearly suggested by the COBE-DMR4
results \citep{DMR4}.

Although the choice of $\Shalf$ was \textit{a posteriori} for the analysis
of the first year of data from WMAP, it is not for the present analysis of
three and five year WMAP data. In their three and five year data releases
the WMAP team has improved the calibration of the CMB maps and their
understanding of systematic issues. Thus, there was the possibility that
the lack of correlation would go away, but --- as demonstrated below --- it
persists.

\begin{table*}
  \begin{minipage}{5in}
    \caption{The $\mathcal{C}_\ell$ calculated from $\mathcal{C} (\theta)$
      for the various data maps.  The WMAP (pseudo and reported MLE) and
      best-fit theory $C_\ell$ are included for reference in the bottom five
      rows.}
    \label{tab:Cl}
    \begin{tabular}{lcccccc} \hline
      Data & \Shalf & $P (\Shalf)$ & $6\mathcal{C}_2/2\pi$ &
      $12\mathcal{C}_3/2\pi$ & $20\mathcal{C}_4/2\pi$ &
      $30\mathcal{C}_5/2\pi$ \\ 
      Source & $(\mu\rmn{K})^4$ & (per cent) & $(\mu\rmn{K})^2$ &
      $(\mu\rmn{K})^2$ & $(\mu\rmn{K})^2$ & $(\mu\rmn{K})^2$ \\
      \hline
      V3 (kp0, DQ) & $1288$ & $0.04$ & $77$ & $410$ & $762$ & $1254$ \\
      W3 (kp0, DQ) & $1322$ & $0.04$ & $68$ & $450$ & $771$ & $1302$ \\
      ILC3 (kp0, DQ) & $1026$ & $0.017$ & $128$ & $442$ & $762$ & $1180$ \\
      ILC3 (kp0), $\mathcal{C} (>60\degr)=0 $ & $0$ & --- & $84$ & $394$ &
      $875$ & $1135$ \\
     ILC3 (full, DQ) & $8413$ & $4.9$ & $239$ & $1051$ & $756$ & $1588$ \\
      \hline
      V5 (KQ75) & $1346$ & $0.042$ & $60$ & $339$ & $745$ & $1248$ \\
      W5 (KQ75) & $1330$ & $0.038$ & $47$ & $379$ & $752$ & $1287$ \\
      V5 (KQ75, DQ) & $1304$ & $0.037$ & $77$ & $340$ & $746$ & $1249$ \\
      W5 (KQ75, DQ) & $1284$ & $0.034$ & $59$ & $379$ & $753$ & $1289$ \\
      ILC5 (KQ75) & $1146$ & $0.025$ & $81$ & $320$ & $769$ & $1156$ \\
      ILC5 (KQ75, DQ) & $1152$ & $0.025$ & $95$ & $320$ & $768$ & $1158$ \\
      ILC5 (full, DQ) & $8583$ & $5.1$ & $253$ & $1052$ & $730$ & $1590$ \\
      \hline
      WMAP3 pseudo-$C_\ell$ & $2093$ & $0.18$ & $120$ & $602$ & $701$ & $1346$ \\
      WMAP3 MLE $C_\ell$ & $8334$ & $4.2$ & $211$ & $1041$ & $731$ & $1521$ \\
      Theory3 $C_\ell$ & $52857$ & $43$ & $1250$ & $1143$ & $1051$ & $981$ \\
      WMAP5 $C_\ell$ & $8833$ & $4.6$ & $213$ & $1039$ & $674$ & $1527$ \\
      Theory5 $C_\ell$ & $49096$ & $41$ & $1207$ & $1114$ & $1031$ & $968$ \\
      \hline
    \end{tabular}
  \end{minipage}
\end{table*}

The calculation of \Shalf\ by direct use of~(\ref{eq:S12}) is susceptible
to noise in $\mathcal{C}(\theta)$.  To avoid this we calculate \Shalf\
directly from $\mathcal{C}_\ell$ as
\begin{equation}
  \Shalf = \frac1{(4\pi)^2}\sum_{\ell,\ell'} (2\ell+1)(2\ell'+1)
  \mathcal{C}_\ell \mathcal{I}_{\ell, \ell'}\left(1/2\right) \mathcal{C}_{\ell'}.
\end{equation}
The calculation of $\mathcal{I}_{\ell,\ell'}(x)$ is described in
Appendix~\ref{app:Illp}.  The $\mathcal{C}_\ell$ smooth over
$\mathcal{C}(\theta)$ as defined in Eq.~(\ref{eq:calCl}).

We can use \Shalf\ to characterize the likelihood of the observed correlation
function.  For the COBE-DMR data \citep{DMR4}, there are relatively large
error bars on $\mathcal{C}(\theta)$, which are consistent with a wide range of
\Shalf\ ranging from under $1000\unit{(\mu{K})^4}$ to approximately
$6000\unit{(\mu{K})^4}$.  But to understand the significance of these values,
we must compare them to those obtained from random realizations of the sky in
the concordance $\Lambda$CDM model with the best-fit parameters.  For this
comparison, we generated maps based on the WMAP five-year $\Lambda$CDM MCMC
parameter chain.  There are 20,401 sets of parameters in this chain.  We
computed the $C_\ell$ for these parameter sets using
\texttt{CAMB}~\citep{CAMB}.  For the $C_\ell$ corresponding to each set of
parameters, we generated a number of random maps (i.e.\ maps with $a_{\ell m}$ 
drawn from Gaussian distributions with zero mean and variance $C_\ell$) based on the weight
assigned to each WMAP MCMC parameter set.  This produced a total of 99,997
maps at \texttt{NSIDE=64}.  From the distribution of \Shalf\ values generated
we calculated the probability ($p$-value) of randomly attaining a \Shalf\ as
low as those we found.  For COBE-DMR the maximum value of \Shalf\ of
$6000\unit{(\mu{K})^4}$ corresponds to a 3 per cent chance of obtaining this little
angular correlation in a random realization of the concordance model.

\section{Basic results}

Table \ref{tab:Cl} lists (columns 2 and 3) the value of \Shalf\ and its
$p$-value among the sample of 99,997 WMAP MCMC maps for the three-year and
five-year maps related to those plotted in Fig.~\ref{fig:ctheta}.  These
include the three and five-year V band (V3 and V5) and W band (W3 and W5)
cut-sky maps.  Also the ILC map in three-year and five-year versions, both
full and cut sky.  We use the kp0 cut for three-year maps and the KQ75 cut
in the case of the five-year maps.  The five-year cut-sky maps are
presented both as measured and corrected for the contribution of the
Doppler quadrupole (DQ), see e.g.~\citet{Schwarz2004}.  (All maps are
corrected by the WMAP team for the Doppler dipole.)

The Legendre transform of $\mathcal{C} (\theta)$ gives us $\mathcal{C}_\ell$,
and these values are also listed in Table~\ref{tab:Cl} (columns 4--7) for
$\ell=2\hbox{--}5$.  Also included in the table, in the bottom five rows, are
the $\ell=2\hbox{--}5$ values of the WMAP three-year
pseudo-$\mathcal{C}_\ell$, the WMAP three-year $\mathcal{C}_\ell$ as extracted
by the WMAP team using a maximum likelihood analysis (the reported values of
the $\mathcal{C}_\ell$), the reported five-year values of the
$\mathcal{C}_\ell$, and the theoretical $\mathcal{C}_\ell$ computed using the
best-fit parameter values as reported both in the three-year and the five-year
WMAP analysis.  Finally and importantly, the table also shows the values of
\Shalf\ and their $p$-values computed from the Legendre transform of these
angular power spectra.

The three cut-sky maps, V, W, and ILC, whether three-year or five-year, are
all in good agreement with each other.  They all have very low values of
\Shalf\ --- almost two orders-of-magnitude below the predictions of the
theory.  In both the three and five-year ILC outside the Galaxy, the
probability that such low values could happen by chance is
extraordinarily low --- only $0.025$ per cent.  This low probability is entirely
consistent with the original finding of COBE-DMR \citep{DMR4} described
above, however the error bars on $\mathcal{C}(\theta)$ (and hence on
$S_{1/2}$) have declined substantially, with the WMAP value of \Shalf\
being at the absolute lowest end of what was consistent with COBE, and with
much smaller error bars.  This dramatic decline in the error bars, while
honing in on the very low end of the COBE-DMR range, is exactly what one
would expect from the absence of large-angle correlations in the
CMB sky, and not at all what one would expect if the low \Shalf\ in
COBE-DMR (and in WMAP) was merely a statistical fluctuation in the
measurement.

We also consider the case where there is exactly zero large scale angular
correlations.  That is, we set $\mathcal{C}(\theta\ge60\degr)=0$ and extract
the $\mathcal{C}_\ell$ as a Legendre transform\footnote{We note that setting
$\mathcal{C}(\theta\ge60\degr)=0$ introduces a small monopole into the power
spectrum.  This can be corrected by subtracting it out, changing the
$\theta_c=60\degr$ to a value such that $\int_0^{\theta_c}
\mathcal{C}(\theta) \sin\theta \dderiv\theta=0$, etc.  Without an
underlying theory to describe this case it isn't clear how to best correct
for this monopole.  Regardless, the $\mathcal{C}_\ell$ we extract are not
very sensitive to the method we use for removing the monopole.}
for the ILC (kp0) map. This ``theory'' produces low-$\ell$
$\mathcal{C}_\ell$ of approximately the same value as for the
$\mathcal{C}(\theta)$ from the cut-sky maps. On the one hand this is
consistent with the statement that there is little correlation on large
angular scales and thus the $\mathcal{C}_\ell$ for low-$\ell$ are dominated by
small angular scales.  On the other hand, this shows that the data is
consistent with there being \textit{no} correlations on large angular scales.

We note that it is difficult to enforce $\mathcal{C}(\theta)=0$ in the context
of a statistically isotropic model.  Even if a model were found that
predicted the observed $C_\ell$ as the expected means of the 
$\left\vert a_{\ell m}\right\vert^2$
(as in equation \ref{eq:SI}), any actual realization of the Universe
would produce $C_\ell$ that were  substantially different.  
Indeed, we have found that $>97$ per cent of realizations of 
such a Universe would have values of \Shalf\ greater than the
observed value (see section~\ref{sec:lowl}).

The results from the full-sky ILC map, also show low values of \Shalf;
however, with less remarkable $p$-values of 5 per cent in contrast to
$0.02\hbox{--}0.04$ per cent for the various cut-sky maps.  Similarly, the full
sky ILC maps have larger quadrupoles than the cut-sky maps (though still
lower than expected from theory), and octopoles consistent with theory.
These full-sky maps are in good agreement with the WMAP-reported MLE
$C_\ell$.  Meanwhile the pseudo-$C_\ell$ based on the kp2 sky-cut (which
cuts less of the sky than the kp0 cut) are intermediate between the kp0
cut-sky map results and the full-sky results.
 
Thus the full-sky results seem inconsistent with cut-sky results and they
appear inconsistent in a manner that implies that \textit{most of the
large-angle correlations in reconstructed sky maps are inside the part of
the sky that is contaminated by the Galaxy}.

\section{Missing power or unfortunate alignment?}
\label{sec:rotate_ILC}

An important question to consider is whether the extremely low large-angle
correlations in the cut-sky WMAP maps are a general result of cutting the
maps or is specific to the orientation of the cut.  That is, should we
expect a loss of large-angle correlations in a cut-sky map or is the
alignment of the cut with the Galaxy important. To address this question
the full-sky five-year ILC map was randomly rotated (that is, set its north
pole in a random direction and draw its azimuthal angle from a uniform
distribution) 300,000 times. For each random rotation we masked the map
with the Galactic KQ75 mask, which is now, therefore, randomly oriented
relative to the original map and re-computed the quadrupole, octopole and
\Shalf\ statistic.

The analysis shows that the true cut-sky quadrupole and octopole are not
terribly unusual compared to those inferred from the rotated-then-cut (RTC)
maps.  In 7.6 per cent of these RTC maps the quadrupole is smaller than that of
the ILC with the originally placed cut, while 2.5 per cent have a smaller
octopole.  Therefore, if we looked at the quadrupole and octopole alone we
would conclude that an arbitrarily oriented mask is only moderately
unlikely to produce low-$\ell$ power in the cut-sky ILC.  Conversely, the
particular alignment of the Galaxy with the part of the sky on which
the low-$\ell$ power is concentrated is only moderately important.

On the other hand, in the RTC maps $\Shalf=11900\unit{(\mu K)^4}$ with
variance $7300\unit{(\mu K)^4}$, a very high value relative to the original
cut ILC map ($1152\unit{(\mu K)^4}$, see Table \ref{tab:Cl}). Only 2 per
cent of these rotated maps have \Shalf\ lower than the ILC with the
original cut. (Recall that $\Shalf\simeq 8583\unit{(\mu K)^4}$ for the
full-sky ILC is already low, with a $p$-value of only about 5 per cent.)
Thus it is quite unlikely for an arbitrary cut to suppress the large-angle
correlations to the extent observed in the cut-sky ILC map.  Conversely, it
is quite likely that the observed absence of large-angle correlations
outside the KQ75 cut is due to the alignment of the Galaxy with the regions
on the sky where such correlations are maximized.  This result is in good
agreement with the result from \citet{Hajian:2007pi} that the little
correlation above $60\degr$ stems from two specific regions within the
Galactic cut covering just 9 per cent of the sky.

\begin{figure}
  \includegraphics[angle=-90,width=3.5in]{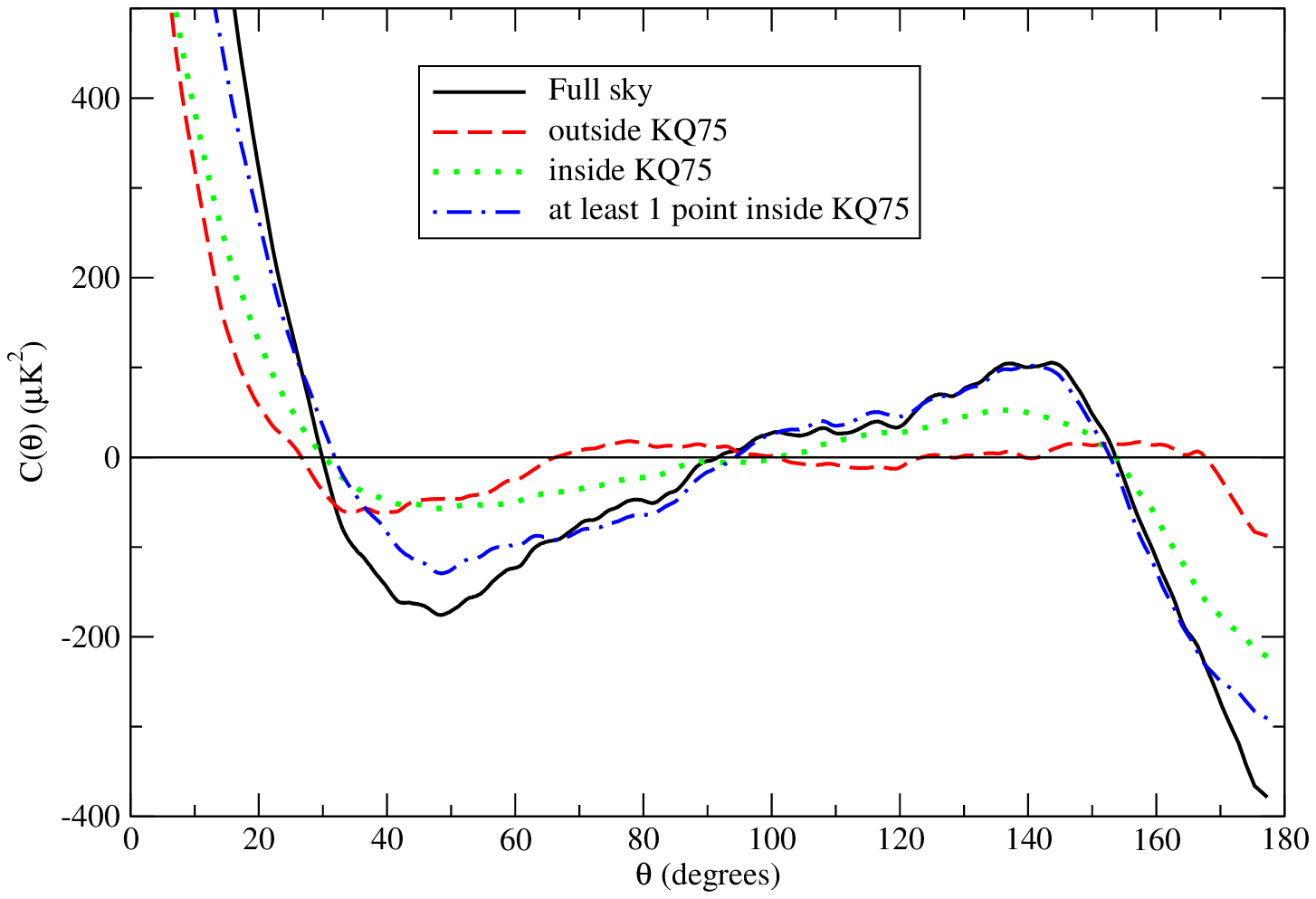}
  \caption{The two-point angular correlation function from the WMAP 5 year
    results.  Plotted are $\mathcal{C} (\theta)$ for the ILC calculated
    separately the part of the sky outside the KQ75 cut (dashed line),
    inside the KQ75 cut (dotted line), and on the part of the sky with at
    least on point inside the KQ75 cut (dotted-dashed line).  For better
    comparison to the full-sky $\mathcal{C} (\theta)$ (solid line), the
    partial-sky $\mathcal{C} (\theta)$ have been scaled by the fraction of
    the sky over which they are calculated.  }
  \label{fig:cutcutvsskysky}
\end{figure}

It appears that our microwave background sky has anomalously low angular
correlations everywhere outside the Galactic mask, but not within.  In
Fig.~\ref{fig:cutcutvsskysky} we plot $\mathcal{C} (\theta)$ for the WMAP 5
year ILC map calculated separately on the part of the sky outside the KQ75
cut, inside the KQ75 cut, and on the part of the sky with at least one point
inside the KQ75 cut.  For better comparison to the full-sky $\mathcal{C}
(\theta)$ (also plotted), the partial-sky $\mathcal{C} (\theta)$ have been
scaled by the fraction of the sky over which they are calculated.  This shows
that the full-sky $\mathcal{C} (\theta)$ is very close to $\mathcal{C}
(\theta)$ calculated from the masked region. Meanwhile
$\mathcal{C} (\theta)$ calculated outside the Galactic mask is similar to
neither, and much closer to zero in magnitude.

Figure \ref{fig:cutcutvsskysky} shows two other interesting peculiarities of
the measured angular correlation function.  First, the full-sky $\mathcal{C}
(\theta)$ is particularly closely mimicked by the $\mathcal{C} (\theta)$
computed so that at least one of the points is inside the masked region; 
the agreement between the two at large angles (above approximately 
$60\degr$) is nearly perfect. Moreover, all four correlation functions shown in 
the figure vanish at nearly the same angle, $\theta\sim 90\degr$. While at 
this time we do not understand the origin or significance of these two 
effects, we wish to point them out because it is possible that they will 
be useful for successful theoretical or systematic explanations of the 
vanishing correlation function.

The evidence we present strongly suggests that the full-sky ILC map does
not represent a statistically isotropic microwave sky.  If the region outside
the cut is a reliable representation of the CMB then we should focus on the
angular correlation for cut skies.  As shown above this leads to a $p$-value
of 0.025 per cent for the standard $\Lambda$CDM model (see Table \ref{tab:Cl}).
Furthermore, {\em the WMAP reported MLE $\mathcal{C}_\ell$, which assumes
  Gaussianity and statistical isotropy in their calculation, are in good
  agreement with the full-sky $\mathcal{C}_\ell$ and $\mathcal{C}(\theta)$,
  but not with their cut-sky equivalents, whereas, the cut-sky
  $\mathcal{C}_\ell$ and $\mathcal{C}(\theta)$ are in good agreement with
  the pseudo-$C_\ell$ (see Table
  \ref{tab:Cl} and figure \ref{fig:ctheta}). This casts doubt on the validity
  of the reported low-$\ell$ $C_\ell$.}  

\section{So, is the large scale CMB anomalous or not?}

It has been suggested that there is nothing particularly anomalous about
the large-angle CMB \citep{Efstathiou2003,Gaztanaga2003,Slosar2004}.
The argument goes something like this: (a) the
two-point angular correlation function $\mathcal{C} (\theta)$ and the
angular power spectrum $C_{\ell}$ contain the same information; (b) not
only does theory tell us that the $C_{\ell}$ are the ``relevant''
variables, but, since they are discrete and finite in number, we can apply
standard statistical techniques to compare observations with theoretical
predictions;  when we do so, we find that only $C_2$ is far below the
expected value, but still at a level that happens by chance 10 per cent of the
time in the concordance model.

We have already pointed out above that (a) contains two fallacies.
First, $\mathcal{C} (\theta)$ and $C_{\ell}$ are equivalent only on a full sky.
Second, formal equivalence is not the only question, signals are often 
more visible in one representation of the data than in a different, though
formally equivalent one.  If this were not so, then there would be no need
for Fourier analysis.  Nor would we need to perform great music --- we could
simply read the score.

Point (b) is correct if the Inflationary Cold Dark Matter ($\Lambda$CDM)
cosmological model is true. This model tells us that the spherical harmonic
coefficients $a_{\ell m}$ are independent Gaussian random variables, and
that the sky is statistically isotropic, so that the off-diagonal
covariances of the estimator $C_\ell$ (as defined in Eq.~\ref{eq:Cl})
vanishes in linear theory.  This vastly simplifies statistical analysis of
the CMB in the context of $\Lambda$CDM.  However, in advancing the case for
a particular cosmological model we are required not just to determine the
best-fit parameters of the model but to test the assumptions and other
predictions of the model.  This includes the prediction of statistical
isotropy, and consequently that the $a_{\ell m}$ are independent of one
another.

There is already considerable evidence that if one analyzes the full-sky
ILC map that one finds difficult-to-explain deviations from statistical
isotropy, such as the alignment of the octopole and quadrupole with each
other and with the geometry of the Solar System (for example
\cite{deOliveira2004, Eriksen_asym, Schwarz2004, Land2005a}).  These
analyses require full-sky data for any statistical power, and so, in
particular, might be explained by Galactic foreground contamination
(though it would be an odd coincidence for Galactic contamination to cause
alignment with the Solar System).  The calculation of $\mathcal{C}
(\theta)$, as we have seen, can be done on a cut sky.

\subsection{Are just the low-$\ell$ $C_\ell$ incorrect?}
\label{sec:lowl}

Perhaps the standard assumptions of statistical isotropy and Gaussianity 
are correct and only the low-$\ell$ $C_\ell$ are incorrectly 
predicted by the standard model.  If this were the
case then the $a_{\ell m}$ could still be Gaussian random variables and
some new physics would be needed to explain the low-$\ell$ $C_\ell$,
e.g., by giving up the scale invariance of the primordial power
spectrum, which could be caused by a feature in the inflationary 
potential.

To study this possibility we replaced $C_2$ through $C_{20}$ in the best
fit $\Lambda$CDM model with the values extracted from the cut-sky ILC
five-year map.  From these $C_\ell$ 200,000 random maps were created,
masked, and \Shalf\ computed.  Under the assumptions of Gaussianity and
statistical isotropy of these $C_\ell$ only 3 per cent of the generated maps had
\Shalf\ less than $1152\unit{(\mu K)^4}$ (the cut-sky ILC5 value from
Table~\ref{tab:Cl}).  Thus even if the $C_\ell$ are set to the specific
values that produce such a low \Shalf\ a Gaussian random, statistically
isotropic realization is 97 per cent unlikely to produce the observed sky.
Again this shows that either (i) the low-$\ell$ $C_\ell$ are correlated,
thus breaking statistical isotropy, or (ii) our Universe is a 97 per cent
unlikely realization of an alternative model that deviates from the
standard one as required to produce the low-$\ell$ $C_\ell$.

\subsection{Statistics of $C(\theta)$}

Once we have decided to calculate $\mathcal{C} (\theta)$, 
we are forced to ask how best to analyze it statistically.  
One option would be to  compare the $\mathcal{C} (\theta)$ inferred
from a particular map to the  $C (\theta)$ one expects from theory.
Thus one would define
\begin{equation}
  C^{\mathrm{th}} (\theta) \equiv  \frac1{4\pi} \sum_{\ell=0}^\infty
  (2\ell+1) C_\ell^{\mathrm{th}} P_\ell (\cos\theta), 
\end{equation}
for a particular set of parameters (say the best-fit values) of
the concordance model. This is what is plotted in figure~\ref{fig:ctheta}
as the $C (\theta)$ of the best-fit $\Lambda$CDM model.

One would next define some functional norm and compute
\begin{equation}
  \label{eqn:theorynorm}
  N^{\mathrm{obs-\Lambda{CDM}}} \equiv \vert\vert \mathcal{C}(\theta) 
  - C^{\mathrm{th}}(\theta)\vert\vert
\end{equation}
where we imply a suitable average over a range of $\theta$ on the right-hand
side.  This norm could serve as a statistic to compare the two-point
correlation function inferred from the data, or some subset of the data to the
theory. The shaded band around $C^{\mathrm{th}} (\theta)$ in
figure~\ref{fig:ctheta} (cosmic variance)
reflects this notion that one somehow expects the inferred 
$\mathcal{C}(\theta)$ to lie inside this band.

The statistic originally suggested by the WMAP team for comparing
observations of large-angle correlations to theory, \Shalf, does not fall
into the above class of statistics.  This is because it captures
that \textit{what is strange about the inferred angular correlation
  function} $\mathcal{C} (\theta)$ \textit{is not that it is different than
  theory for} $\theta\ga60\degr$, \textit{but rather that it is so close
  to zero.}  Thus \Shalf\ is designed to test an alternative simple
hypothesis --- that there are no correlations above $60\degr$.  In the
language of equation (\ref{eqn:theorynorm}) \Shalf\ is in the class of
statistics
\begin{equation}
  N^{\mathrm{obs-zero}}  \equiv  \vert\vert \mathcal{C}(\theta) -
  0\vert\vert .
\end{equation} 

There is another lesson to be learned from the preceding results. Cosmological
inflation predicts that there are fluctuations on all scales, whereas many
alternative models of structure formation, like cosmic defects, would predict
the absence of fluctuations on super-horizon scales. By looking at scales
above $1$ degree on the sky the inflationary prediction is tested at the time
of photon decoupling, and by looking at the largest angular scales, we can
test it in the more recent Universe since the physical Hubble scale $r_H(z) =
1/H(z)$ is observed at the angle $\theta = r_H(z)/d_{\rm a}(z)$ and
angular distance $d_{\rm a}(z) = [1/(1+z)] \int_0^z [1/H(z')] \dderiv z'$.  
For the best-fit parameters of the concordance model, the lack of 
correlations at larger $60$ degrees means, that scales that crossed into 
the Hubble radius below a redshift $\approx 1.5$ are uncorrelated.
 
Instead of \Shalf, \citet{Hajian:2007pi} advocates the use of a
covariance-weighted integral over $\mathcal{C} (\theta)$,
\begin{equation}
  \label{eqn:HajianA}
  A(x) \equiv \int_{-1}^x \dderiv(\cos\theta)\int_{-1}^x \dderiv (\cos\theta') 
  \mathcal{C}(\theta)F^{-1}(\theta,\theta')\mathcal{C}(\theta'),
\end{equation}
where
\begin{equation}
  F(\theta,\theta') \equiv 
  \left< (C(\theta)-\langle C(\theta)\rangle)
    (C(\theta')-\langle C(\theta')\rangle)\right>,
\end{equation}
and $\left<\cdots\right>$ represents an ensemble average, i.e. an average
over realizations of the underlying $\Lambda$CDM model.  As Hajian notes,
$A(1/2)=\Shalf$ in the limit of uncorrelated $\mathcal{C} (\theta)$.
However, just because $C (\theta)$ and $C (\theta')$
are correlated in the standard theory does not make $A(1/2)$ a more correct
statistic than $\Shalf$.  For tests against the standard theory the $A(x)$
statistic provides \textit{another} statistic; one that accounts  for
the theory correlations.  However, as discussed above, it has repeatedly
been shown that there are correlations among the low-$\ell$ multipole
moments (and multipole vectors) of the full sky that are not consistent
with the standard theory. In this case it is not possible to compute
$F(\theta,\theta')$ because the ensemble over which to average is unknown.
Therefore, while it is somewhat reassuring that by using $A(x)$
\citet{Hajian:2007pi} confirms our earlier result \citep{wmap123} showing
$\mathcal{C} (\theta)$ for cut skies violates the fundamental model
assumption of statistical isotropy, it is not clear that any strong
inference should be drawn from differences in statistical significance
between results for $A(1/2)$ and $\Shalf$.

\begin{table*}
  \begin{minipage}{4.5in}
    \caption{\Shalf\ (in $(\mu\mathrm{K})^4$) obtained by minimizing with
      respect to $C_\ell$. We show the statistic for the best-fit theory
      and WMAP, as a function of the cutoff multipole $\lmaxtune$ (the
      minimization has been performed by varying all $\ell$ in the range
      $2\leq \ell\leq \lmaxtune$ and fixing $\ell>\lmaxtune$).  Also shown
      is the 95 per cent confidence region of the minimized \Shalf\ derived from
      chain 1 of the WMAP MCMC parameter fit.  In the bottom row, we remind
      the reader that the measured value of \Shalf\ outside the cut is
      $1152\unit{(\mu K)^4}$ (see Table~\ref{tab:Cl} for more
      details).
}
    \label{tab:minS12}
    \begin{tabular}{lccccccc}
      \hline
      $C_\ell$ & \multicolumn{7}{c}{Maximum tuned multipole,  $\lmaxtune$} \\ \cline{2-8}
      Source & $2$ & $3$ & $4$ & $5$ & $6$ & $7$ & $8$ \\ \hline
      Theory & $7624$ & $922$ & $118$ & $23$ & $7$ & $3$ & $0.7$ \\
      Theory 95 per cent & $6100$--$12300$ & $750$--$1500$ & $100$--$200$ &
      $20$--$40$ & $7$--$14$ & $3$--$6$ & $1$--$3$ \\
      WMAP & $8290$ & $2530$ & $2280$ & $800$ & $350$ & $150$ & $130$ \\
      \hline
	ILC5 (KQ75) & \multicolumn{7}{c}{$1152$}\\
	\hline
    \end{tabular}
  \end{minipage}
\end{table*}

The $A(x)$ statistic suggested by \citet{Hajian:2007pi} and the MLE estimator
for the $C_\ell$ are examples of optimal statistics.  These statistics have
minimum variance \textit{for a specific theory}.  In both these cases the
assumptions of Gaussianity and statistical isotropy are employed.  Once a
theory is established these statistics make optimal use of the available data
to extract the most precise possible values of model parameters or of 
values of summary properties of the data, for example of the $C_\ell$.  However, when
testing the validity of a theory they only provide \textit{another} statistic
and may not provide the best test of the assumptions of that theory.  In the
work presented here, we implicitly assume a flat weighting of the pixel
temperatures in computing $\mathcal{C}(\theta)$; see
Eq.~(\ref{eq:calCthetaavg}). Furthermore, when assessing the lowness of
$\mathcal{C}(\theta)$ at large scales, we do not rely on any particular underlying theory and
assume a flat weighting implicit in the definition of our statistic \Shalf.  We find that if
we assume Gaussianity and statistical isotropy (through use of the MLE
$C_\ell$, see Table~\ref{tab:Cl}) then the standard model has a $p$-value of
5 per cent.  However, if we do \textit{not} make these assumptions then 
the standard model only has a $p$-value of 0.025 per cent.  Without a much 
more detailed analysis, it seems to us that a flat weighting is more robust 
against incorrect assumptions about the actual statistical distribution 
than an optimal weighting. In order
to definitively answer that question one would need to analyze the higher
($n$-point) correlation functions at large angular scales, which is beyond the
scope of this work.

Finally, we again emphasize that what is anomalous about the observed
large-angle correlations is \textit{not} how poorly they match the theory,
but rather how well they agree with the very simple alternative
phenomenological hypothesis that there are no large-angle correlations ---
$\mathcal{C}(\theta> 60\degr)=0$.  The construction of
$N^{\mathrm{obs-\Lambda{CDM}}}$, might indeed benefit from an attempt to
remove expected correlations through $F(\theta,\theta')$, as in
\cite{Hajian:2007pi}; however, the theoretical model against which
$N^{\mathrm{obs-zero}}$ compares the observations has
$F(\theta,\theta')\equiv0$ for the relevant $\theta$.

\subsection{Minimizing \Shalf}\label{sec:minS12}

Once we have understood that what is anomalous about $\mathcal{C} (\theta)$
is how close it is to zero, we can understand that what is strange about
the low-$\ell$ $C_\ell$ is not just how low $C_2$ is, but also how the
various $C_\ell$ are correlated with each other.
 
We now probe the sensitivity of \Shalf\ to ranges of $\ell$.  Given that
small angles can affect low-$\ell$ results, it is also the case that higher
$\ell$ can affect the larger angles.  One way to see this is to determine
how the $C_\ell$ for low-$\ell$ must be adjusted to attain a 
low \Shalf.
This is not done by setting some range of $C_\ell$ 
to zero.  Instead, given
a set of $C_\ell$ for $\ell>\lmaxtune$ we can find the values 
of $C_\ell$
for $2\le\ell\le \lmaxtune$ that minimize \Shalf\, by regarding \Shalf\
as a function of $C_\ell$ using equations 
(\ref{eq:Ctheta}) and (\ref{eq:S12}).

We consider two sets of $C_\ell$, the first from the $\Lambda$CDM theory, the
second as reported by WMAP.  Table~\ref{tab:minS12} shows the minimum \Shalf\
we find for each value of $\lmaxtune$.  In the table, we provide the values
for the best-fit $\Lambda$CDM model and the reported WMAP $C_\ell$.  We also
provide the 95 per cent confidence ranges based on the WMAP MCMC parameter set chain
where the minimum \Shalf\ was calculated independently for each model.

To attain $\Shalf\le1152\unit{(\mu{K})^4}$ (the value found in the
masked ILC map, see Table~\ref{tab:Cl}) from the best-fit theory requires
tuning both $C_2$ and $C_3$.  Thus \textit{even the theory requires more
  fine-tuning than just the quadrupole to be low in order to be consistent with
  observations}.  The minimum in Table~\ref{tab:Cl} was attained for
$6C_2/2\pi=149\unit{(\mu{K})^2}$ and
$12C_3/2\pi=473\unit{(\mu{K})^2}$. In general for the theory we need to
tune at least up to $\lmaxtune=3$ and can almost always find a low \Shalf\
if we tune up to $\lmaxtune=4$.

For the WMAP $C_\ell$ even more tuning is required.  Note that
the WMAP $C_2$ is already approximately tuned to produce the
minimum \Shalf\ given the rest of the $C_\ell$ for $\ell>2$ (that
is, from Table~\ref{tab:Cl} we note that $8583\unit{(\mu{K})^4} \approx
8290\unit{(\mu{K})^4}$).  To attain the low \Shalf\ to match the cut-sky
ILC requires tuning of values of $C_\ell$ up to $\lmaxtune=5$.

Table~\ref{tab:minS12} further shows that the minimum \Shalf, that can be
achieved by optimizing the low-$\ell$ $C_\ell$ fall off
much more slowly in the WMAP $C_\ell$ than in the theory.  By
$\lmaxtune=8$ the minimum WMAP \Shalf\ is two orders of magnitude larger
than can be attained from the theory.  This strongly suggests that
important correlations exist in the data for $\ell\ge8$ that do not exist
in the theory.  These correlations cannot be canceled by tuning the lower
$\ell$ behavior.

Therefore, we conclude that a given behavior of $C (\theta)$ on
large scales is \textit{not} in unique relation to a behavior of
$C_\ell$ at low-$\ell$. The former quantity receives significant
contributions from $C_\ell$ at high $\ell$ as well; the converse
is also true.  Given the extremely puzzling near-vanishing power in
$\mathcal{C}(\theta>60\degr)$, and given that the quadrupole and octopole
are \textit{not} unusually low (as shown in e.g.\ \citet{O'Dwyer2004}), we
argue that any theoretical or observational explanation of the ``low power
at large scales'' should concentrate on the quantity $\mathcal{C}
(\theta)$.

\section{Conclusions}\label{sec:conclude}

In this paper we have studied the angular correlation function in WMAP
three- and five-year maps. We have clarified the relation between various 
definitions of
the angular correlation function, and revisited our previous calculation
from \citet{wmap123} in more detail. We confirmed that power on large
angular scales --- greater than about 60 degrees --- is anomalously low, at
99.975 per cent CL (see Table \ref{tab:Cl}).  The measured angular correlation
function thus disagrees with the $\Lambda$CDM theory, but, more
significantly, it is consistent with a simple phenomenological ``theory''
--- $\mathcal{C} (\theta\ge60\degr)\equiv 0$.  The significance of this
disagreement (as measured by the probability of the value of $S_{1/2}$) has
now increased by a factor of over 100 since it was first observed in the
COBE-DMR four-year analysis.

We have shown that the cut-sky and full-sky large-scale angular
correlations differ (see Table~\ref{tab:Cl} and Figure~\ref{fig:ctheta}),
though the source of these discrepancies remains unknown.  This shows that
either the Universe is \textit{not} statistically isotropic on large
angular scales or that correlations are introduced in reconstructing the
full sky from the observations.  We have shown that even given the
unusually small full-sky angular correlations (95 per cent unlikely) an unusual
alignment of the Galaxy with the CMB (2 per cent of realizations) is required to
explain the lack of correlations outside the Galactic region.  We have
further shown that simply adjusting the theoretical values of the $C_\ell$
does not solve the problem if the sky is representative of a Gaussian
random statistically isotropic process -- the cosmic variance in the
$C_\ell$ is such that less than 3 per cent of all realizations would preserve a
low value of \Shalf\ .

From these results we argue that $\mathcal{C}(\theta)$ is an important
quantity to study along with the usual 
angular power spectrum,
$C_\ell$.  The typical ``rule-of-thumb'' that low-$\ell$ describes large
angular scales is not accurate.  Any theoretical explanations for the ``missing
large-scale power'' should concentrate on explaining the low $\mathcal{C}
(\theta)$, rather than the smallness of the quadrupole and octopole, which are
not nearly as significant \citep{O'Dwyer2004}. As has been
pointed out by \cite{Gordon:2005ai}, \cite{Rakic:2007ve} and \cite{Bunn:2008zd}, any possible
explanation of the multipole alignments that relies on an additive, 
statistically independent contribution to the microwave sky on top of the 
primordial one, increases the significance of the lack of angular correlation.  

The CMB, as measured by WMAP in particular, provides much support for our
current model of the Universe.  It also points the way toward new puzzles
that may affect fundamental physics.  On the largest angular scales the
microwave sky is inconsistent with theoretical expectations.  These
discrepancies between observations and theory remains an open problem.  In
the future, combining the current data with new information, such as new
data from WMAP, observations from the Planck experiment, and polarization
information \citep{Dvorkin} may be key to determining the nature of the
large-scale anomalies.

\section*{Acknowledgements} 

We thank Francesc Ferrer, Lloyd Knox, Eiichiro Komatsu, Aleksandar Raki\'c and
Licia Verde for useful conversations.  Some of the results in this paper have
been derived using the HEALPix \citep{healpix} package and the CAMB software.
We acknowledge the use of the Legacy Archive for Microwave Background Data
Analysis (LAMBDA). Support for LAMBDA is provided by the NASA Office of Space
Science.  CJC and GDS are supported by grants from NASA's Astrophysics Theory
Program and from the US DOE. DH is supported by the DOE OJI grant under
contract DE-FG02-95ER40899, NSF under contract AST-0807564, and NASA under
contract NNX09AC89G.  DJS is supported by grants from DFG.  DJS and GDS thank
the Centro de Ciencias de Benasque for its hospitality.

% For various journals
\newcommand{\apj}{ApJ}
\newcommand{\apjl}{ApJL}
\newcommand{\apjs}{ApJS}
\newcommand{\mnras}{MNRAS}
\newcommand{\prd}{Phys. Rev. D}
\newcommand{\physrev}{Phys. Rev.}
\newcommand{\physrevlett}{Phys. Rev. Lett.}

% To use BIBTeX uncomment these and comment out the input statement below.
\bibliographystyle{mn2e}
\bibliography{angular_correlation}
% To not use BIBTeX comment out the above and uncomment this
%%\input{angular_correlation.bbl}

%%%%%%%%%%
\appendix
%%%%%%%%%%

\section{Integrating products of Legendre polynomials}
\label{app:Illp}

We wish to calculate
\begin{equation}
  \mathcal{I}_{m,n} (x) \equiv \int_{-1}^x P_m (x')P_n (x') \dderiv x'.
  \label{eq:Imn}
\end{equation}
For the special case of $x=1$ this is just the normalization
\begin{equation}
  \mathcal{I}_{m,n} (1) = \frac{2}{2n+1} \delta_{m,n}.
\end{equation}

For a general $x$ we consider two cases.  When $m\ne n$ Legendre's equation
\begin{equation}
  (1-x^2)P_n'' (x) - 2xP_n' (x) +n (n+1)P_n (x)=0
\end{equation}
allows us to write
\begin{eqnarray}
  P_m (x)P_n (x) & = & \frac{\dderiv}{\dderiv x}\left[
    (1-x^2) (P_m'P_n-P_n'P_m)\right] \nonumber \\
  & & {} \bigm/ [n (n+1) - m (m-1)].
\end{eqnarray}
Then using the relation
\begin{equation}
  (1-x^2)P_n' (x) = nP_{n-1} (x) -nxP_n (x)
\end{equation}
we find
\begin{eqnarray}
  \mathcal{I}_{m,n} & = & \bigl\{ mP_n (x)\left[ P_{m-1} (x) - x P_m (x)
  \right] \nonumber \\
  & & \quad {} - n P_m (x)\left[ P_{n-1} (x) - x P_n (x) \right] \bigr\}
  \nonumber \\
  & & {} \Bigm/ \bigl\{n (n+1) - m (m+1)\bigr\} \quad\mbox{[for $m\ne n$]}.
  \label{eq:Imn_offdiag}
\end{eqnarray}

When $m=n$ we integrate~(\ref{eq:Imn}) by parts to get off diagonal
terms~(\ref{eq:Imn_offdiag}) and use the indefinite integral
\begin{equation}
  \int P_n (x) \dderiv x = \frac{1}{2n+1} \left[ P_{n+1} (x) - P_{n-1} (x)\right]
\end{equation}
to derive the recursion relation
\begin{eqnarray}
  \mathcal{I}_{n,n} (x) & = & \bigl\{ \left[P_{n+1}
      (x)-P_{n-1} (x)\right]\left[P_n (x)-P_{n-2} (x)\right] \nonumber \\
    & & \quad {} - (2n-1)\mathcal{I}_{n+1,n-1} (x) \nonumber +
    (2n+1)\mathcal{I}_{n,n-2} (x) \nonumber\\ 
    & & \quad {} + (2n-1)\mathcal{I}_{n-1,n-1} (x) \bigr\}
    \Bigm / \bigl\{ 2n+1 \bigr\}.
    \label{eq:Imn_diag}
\end{eqnarray}
Starting from $\mathcal{I}_{0,0} (x) = x+1$ and $\mathcal{I}_{1,1} (x) =
(x^3+1)/3$ we can calculate all the diagonal terms recursively.

With these two relations~(\ref{eq:Imn_offdiag} and \ref{eq:Imn_diag}) we
can compute and tabulate all required values of $\mathcal{I}_{m,n}$ for
any $x$.

\bsp

\label{lastpage}

\end{document}